\documentclass[superscriptaddress,twocolumn,showpacs,preprintnumbers,amsmath,amssymb]{revtex4}

\usepackage{graphicx}
\usepackage{dcolumn}
\usepackage{bm}

\def\HL{\rm HL}
\def\HXY{\rm HXY}
\def\LXY{\rm LXY}
\def\XYXY{\rm XYXY}
\def\XYF{\rm XYF}
\def\HN{\rm HN}
\def\LN{\rm LN}
\def\stag{\rm stag}
\def\simleq{\mbox{\raisebox{-1.0ex}{$\stackrel{<}{\sim}$}}}
\def\simgeq{\mbox{\raisebox{-1.0ex}{$\stackrel{>}{\sim}$}}}

\begin{document}


\title{Ground State Phase Diagram of $S=1$ XXZ Chains with Uniaxial Single-Ion-Type Anisotropy}

\author{Wei Chen}
\email{wchen20@po-box.mcgill.ca}
\affiliation{%
Department of Chemistry, McGill University, Montreal, PQ, Canada H3A 2K6 \\
}%
\author{Kazuo Hida}
\email{hida@phy.saitama-u.ac.jp}
\affiliation{%
Department of Physics, Saitama University, Saitama, Saitama, Japan 338-8570\\
}%

\author{B. C. {\sc Sanctuary}}
\email{bryan.sanctuary@mcgill.ca}
\affiliation{%
Department of Chemistry, McGill University, Montreal, PQ, Canada H3A 2K6 \\
}%

\date{\today}

\begin{abstract}
One dimensional $S=1$ XXZ chains with uniaxial single-ion-type anisotropy are studied by numerical exact diagonalization of finite size systems. The numerical data are analyzed using conformal field theory, the level spectroscopy, phenomenological renormalization group and finite size scaling method. We thus present the first quantitatively reliable ground state phase diagram of this model. The ground states of this model contain the Haldane phase, large-$D$ phase, N\'{e}el phase, two XY phases and the ferromagnetic phase. There are four different types of transitions between these phases: the Brezinskii-Kosterlitz-Thouless type transitions, the Gaussian type transitions, the Ising type transitions and the first order transitions. The location of these critical lines are accurately determined. 
\end{abstract}

\maketitle

\section{Introduction}
One dimensional antiferromagnetic spin chains have been the subject of recent investigations by numerous groups. A uniform antiferromagnetic Heisenberg chain is known to have a gapless ground state for half-integer-spin. In particular, the exact solution is available for $S=1/2$ chains\cite{bh}. In contrast, for integer-spin\cite{fd}, there is a gap between the first excited state and the ground state. This state is destroyed by various types of perturbations such as single-ion type anisotropy, exchange anisotropy and bond alternation. In this context, the $S=1$ XXZ chain with single ion anisotropy has been studied by many authors from the early stage of the study of Haldane gap problem. In spite of its long history of the study and fundamental importance, however, a quantitatively reliable phase diagram of this model has not yet been published. In the present work, we present the quantitative phase diagram of this model analyzing the exact diagonalization data by various methods including the recently developed level spectroscopy method\cite{nomura} based on conformal field theory and renormalization group.

This paper is organized as follows. In the next section, the model Hamiltonian is defined and the obtained phase diagram is presented. The numerical exact diagonalization results and methods of analysis are explained in \S 3. The final section is devoted to a summary and discussion.

\section{Model Hamiltonian and Ground State Phase Diagram}
The Hamiltonian is given by
\begin{equation}
\label{de}
{\cal H} = \sum_{l=1}^{N}[J({S}_{l}^{x} {S}_{l+1}^{x}+{S}_{l}^{y} {S}_{l+1}^{y})+J_{z}{S}_{l}^{z} {S}_{l+1}^{z}]+D\sum_{l=1}^{N}S_{l}^{z2},
\end{equation}
where $\roarrow {S_{l}}$ is a spin-1 operator. The parameter $D$ represents uniaxial single-ion anisotropy. The periodic boundary condition is assumed unless specifically mentioned. In what follows, we set $J=1$ to fix the energy scale. The ground state phase diagram of this model consists of the Haldance phase, the large-$D$ phase, two XY phases, the ferromagnetic phase and the N\'{e}el phase\cite{md,hj}. Between these phases, various types of phase transitions take place. There is a gapful phase to gapful phase transition (Gaussian transition) between the Haldane phase and large-$D$ phase, gapful-gapless BKT transitions between the XY phase and the Haldane or large-$D$ phase, an Ising transition between the N\'{e}el phase and Haldane phase, a first-order transition between the ferromagnetic phase and the large-$D$ phase or XY phase. The character of the transition between the large-$D$ and N\'{e}el phase  is still unclear although it is likely to be the first order transition.

\begin{figure}[<h>]
\centerline{\includegraphics[width=80mm]{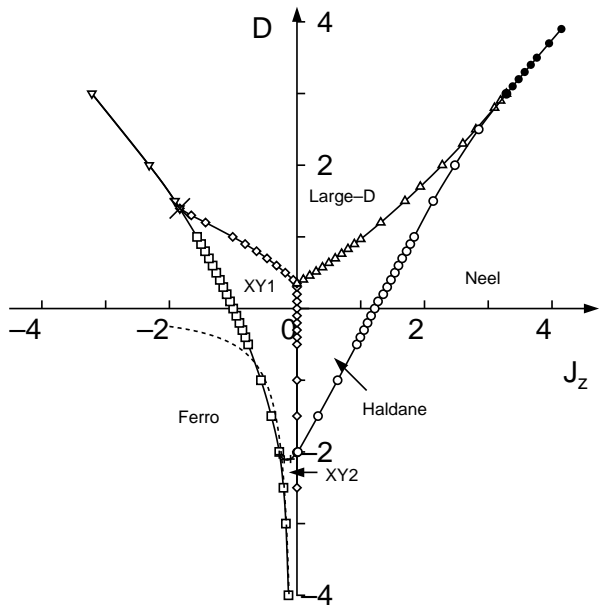}}
\caption{The phase diagram of $S=1$ XXZ chains with uniaxial single-ion-type anisotropy. The solid lines and symbols are the transition lines. The dotted line shows the curve $J_z =-\frac{1}{2\mid D\mid}$ expected from the perturbation calculation for large negative $D$. }
\label{phase}
\end{figure}

Our phase diagram is summarized in Figure \ref{phase}.  For $J_z > 0$, the Haldane-large-$D$ transition line of is shown by the $\triangle$. For large $D$, the ground state becomes a large-$D$ phase, the Haldane phase appears under the large-$D$ phase. With the decrease of $D$, the ground state becomes the N\'{e}el phase. The line with the symbol $\circ$ represents the Haldane-N\'{e}el transition line. For large $D$ and $J_z$, the direct large-$D$-N\'{e}el transition takes place along the line with $\bullet$. For $J_z \le 0$, the XY-Haldane  or XY-large-$D$ BKT transition line is represented by the line with symbol $\diamond$. The XY-ferromagnetic first order transition line is represented by the line with symbol $\square$. The large-$D$-ferromagnetic first order transition line is represented by the line with symbol $\nabla$. The critical line between the two XY phases is denoted by the line with symbol +. The ferromagnetic-XY-large-$D$ tricritical point is represented by $\times$ symbol. In the following, we explain how this phase diagram is obtained by our numerical analysis. 

\section{Numerical Analysis}
\subsection{Haldane-Large-$D$ transition line ($J_{z}>0$ and $D>0$)}
This phase transition is the Gaussian transition. In order to determine the phase boundary with high accuracy, we use the twisted boundary method of Kitazawa and Nomura \cite{ak,kn,kn2}. The Hamiltonian is numerically diagonalized to calculate the two low lying energy levels with the twisted boundary condition ($S^{x}_{N+1}=-S^{x}_1,$ $ S^{y}_{N+1}=-S^{y}_1,$ $ S^{z}_{N+1}=S^{z}_1$) for $N=8, 10, 12, 14$ and 16 by the Lanczos method.

\begin{figure}[<h>]
\centerline{\includegraphics[width=70mm]{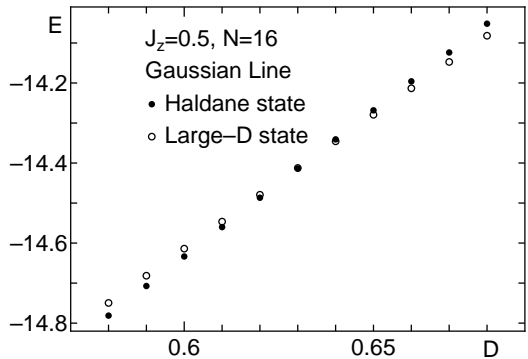}}
\caption{The $D$-dependence of the two lowest energy eigenvalues with twist boundary condition. The energies of the Haldane state and the large-$D$ state are represented by $\bullet$ and $\circ$, respectively, for $N=16$ and $J_z=0.5$. }
\label{gaucross}
\end{figure}

\begin{figure}[<h>]
\centerline{\includegraphics[width=70mm]{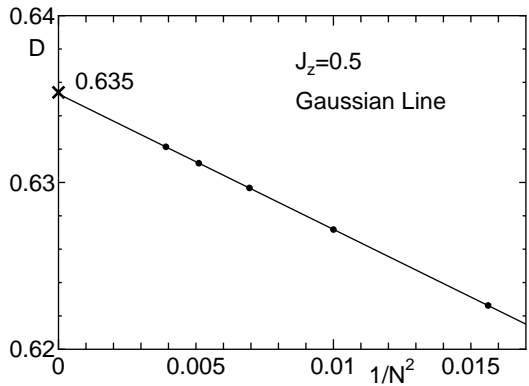}}
\caption{The extrapolation procedure of finite size $D_{c(\HL)}$ for $J_z=0.5$.}\label{expgauline}
\end{figure}

It is known that the ground state is the Haldane phase with a valence bond solid (VBS) structure for small $D$. Under the twisted boundary condition, the eigenvalues of the space inversion $P$ ($\roarrow {S}_{i}\rightarrow \roarrow {S}_{N-i+1}$) and the spin reversal $T$ ($S_{i}^{z}\rightarrow -S_{i}^{z}, S_{i}^{\pm} \rightarrow -S_{i}^{\mp}$) are all equal to $-1$ in this phase\cite{ak,kn,kn2}. As $D$ increases with positive value, a phase transition takes place from the Haldane to the large-$D$ phase for which $P=1$ and $T=1$. The origin of these values of $P$ and $T$ is closely related to the edge spins that characterize the Haldane phase as follows. In the Haldane phase, the two edge spins form a triplet state with positive $P$ and $T$ under the twisted boundary condition. Consequently, those of the whole system, which contains an odd number of singlet pairs, become negative for even $N$. This phenomenon does not take place in the large-$D$ that has no edge spins and these states have positive $P$ and $T$ eigenvalues. Thus we make use of the $P$ and $T$ eigenvalues to distinguish the Haldane phase and the large-$D$ phase with high accuracy. For example, if $J_z$ is fixed, the energies of the two states vary with $D$. For small $D$, the energy of the Haldane state is lower than that of the large-$D$ state. As $D$ increases, the large-$D$ state becomes lower than the Haldane state. The two levels cross at one point $D=D_{c(\HL)}(N)$. This is the finite size Haldane-large-$D$ transition point.  Figure \ref{gaucross} shows the $D$-dependence of the two lowest levels for $N=16$ and $J_z=0.5$. We extrapolate the critical point as $D_{c(\HL)}(N)=D_{c(\HL)}(\infty)+aN^{-2}+bN^{-4}$. Figure \ref{expgauline} shows the extrapolation procedure. The same procedure can be carried out varying $J_z$ with fixed $D$. 


\begin{figure}[<h>]
\centerline{\includegraphics[width=70mm]{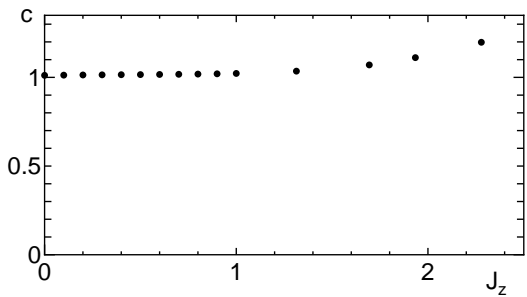}}
\caption{The $J_z$ dependence of the numerically obtained central charge $c$.}
\label{figc}
\end{figure}

In this case the transition line is expected to be described by the conformal field theory with conformal central charge $c=1$. To check this, we estimate the value of $c$. On the critical line, the system is conformally invariant so that a finite size correction to the ground state energy is related to the central charge $c$ and the spin wave velocity $v_{\mbox{s}}$ as follows, \cite{ca,HW,IA}
\begin{equation}
\label{eq1}
\frac{1}{N}E_{\mbox{g}}(N) \cong \varepsilon_{\infty}-\frac{\pi cv_{\mbox{s}}}{6N^{2}},
\end{equation}
\begin{equation}
\label{eq2}
v_{\mbox{s}}=\lim_{N \rightarrow \infty}\frac{N}{2\pi}[E_{k_{1}}(N)-E_{\mbox{g}}(N)].
\end{equation}
Here $E_{\mbox{g}}(N)$ is the ground state energy and $E_{k_{1}}(N)$ is the energy of the excited state with wave number $k_{1}=\frac{2\pi}{N}$ and magnetization $M^{z}(=\displaystyle\sum_{l=1}^{N}S_{l}^{z})=0$.  The ground state has $M^{z}=0$ and $k=0$. Also, $\varepsilon_{\infty}$ is the ground state energy per unit cell in the thermodynamic limit. We calculate $E_{\mbox{g}}(N)$ and $E_{k_{1}}(N)$ by the Lanczos exact diagonalization method under periodic boundary conditions on the transition line. The system sizes are $N=8, 10, 12, 14$ and 16. The size extrapolation is carried out using the formula $c(N)=c+C_{1}N^{-2}+C_{2}N^{-4}$.  The central charge $c$ is close to unity within the range $0 \le J_z \simleq 1$ on the phase boundary as shown in Figure \ref{figc}. Therefore, we expect that the present model can be described by a Gaussian model on the critical line. For $J_z \simgeq 1$, the numerically estimated value of $c$ starts to deviate from unity. Presumably, this is due to the influence of the Haldane-N\'eel Ising critical line that is approaching the large-$D$-Haldane line from below.

\subsection{Haldane-N\'{e}el and large-$D$-N\'eel transition lines ($J_{z}>0$)}

From symmetry consideration, these transition lines are expected to be Ising type transitions. However, for $J_z/J, D/J \rightarrow \infty$, the ground state is determined by the simple classical competition between the ideal large-$D$ state $\mid 00000..>$ for which $E_{\rm g}(N)=0$ and ideal N\'eel states $ \mid 1 -1 1 -1..>$ or $ \mid  -1 1 -1 1..>$ for which $E_{\rm g}(N)=N(D-J_z)$. Therefore a first order transition between these two states is expected. It is not obvious, however, whether quantum fluctuation due to the $J$ term drives this first order transition to a second order transition. 

\begin{figure}[<h>]
\centerline{\includegraphics[width=70mm]{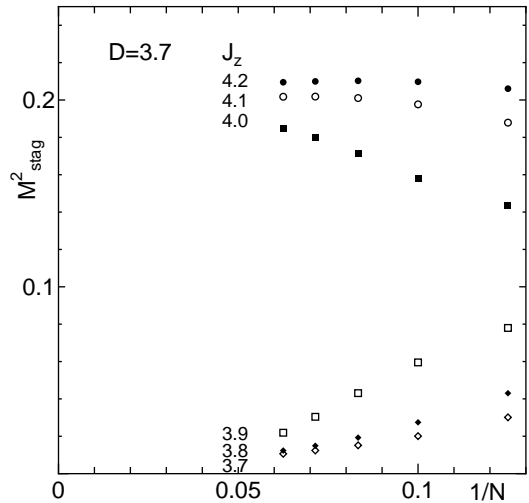}}
\caption{The $N$ dependence of the numerically obtained $M^{2}_{\stag}$ for $D=3.7$ for various values of $J_z$.}
\label{magndep37}
\end{figure}

In order to check this issue, we directly calculate the behavior of the staggered magnetization across the phase boundary. Let us focus on the system size dependence of  $M_{\stag}$. In the N\'eel phase, $M_{\stag}$ should increase with $N$ and tend to a finite value as $N \rightarrow \infty$. On the other hand, it should decrease with $N$ in the large-$D$ phase. The $N$-dependence of $M^{2}_{\stag}$ is plotted against $1/N$ for $D=3.7$ in Figure \ref{magndep37}. The difference of the behavior for $J_z \geq 4.0$ and $J_z \leq 3.9$ is distinct.  Therefore we expect that the phase boundary between the large-$D$ and N\'eel phases is a first order transition line even for finite $D$.

The Haldane-large-$D$ transition line approaches these lines from above and the multicritical point is expected to appear at the crossing point. However, it is not easy to determine the accurate location of the multicritical point. Therefore we discuss the large-$D$-N\'eel and Haldane-N\'eel critical lines as a whole and roughly estimate the position of the multicritical point from the behavior of these two lines and the Haldane-large-$D$ transition line.

\begin{figure}[<h>]
\centerline{\includegraphics[width=70mm]{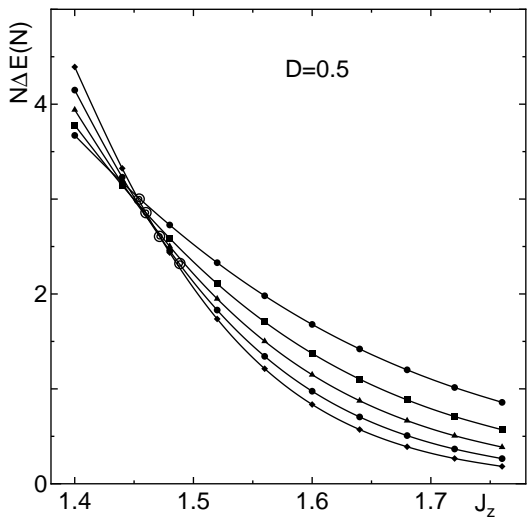}}
\caption{The $J_{z}$-dependence of $N\Delta E(N)$ with $D=0.5$ for $N=$8,10,12,14 and 16. The intersections (double circles) are the finite size critical points.}
\label{hanecross}
\end{figure}

We employ the phenomenological renormalization group (PRG) method to determine these phase transition lines. The Hamiltonian is numerically diagonalized to calculate the lowest energy gap $\Delta E(N)$ in the periodic boundary condition for $N=8,10,12,14$ and 16 using the Lanczos algorithm. The N\'eel state is two fold degenerate in the thermodynamic limit. For finite $N$, this degeneracy is lifted and the energy difference between them gives the smallest gap $\Delta E(N, J_{z}, D)$ which decreases exponentially with $N$. On the other hand, in the Haldene and large-$D$ phases, the energy gap $\Delta E(N)$ remains finite in the thermodynamic limit. Thus the product $N\Delta E(N)$ increases (decreases) with $N$ for $J_{z}<J_{zc(\HN,\LN)}$ ($J_{z}>J_{zc(\HN,\LN)}$) where $J_{zc(\HN)}$ and  $J_{zc(\LN)}$ are the critical value of $J_{z}$ of the Haldane-N\'eel and large-$D$-N\'eel transition, respectively. Furthermore, on the Ising critical line,  the critical exponent for the energy gap is equal to unity. Therefore, the product $N\Delta E(N, J_{z}, D)$ should be size independent for large enough systems in which the contribution from irrelevant operators is negligible. Due to this situation, we can accurately determine the Ising critical point by PRG method. According to the PRG argument, the intersection of $N\Delta E(N)$ for two successive values of $N_1=N$ and $N_2(=N+2)$ defines the finite size critical point $J_{zc(\HN,\LN)}(N_1,N_2)$\cite{mn} as shown in Figure \ref{hanecross}. 

\begin{figure}[<h>]
\centerline{\includegraphics[width=70mm]{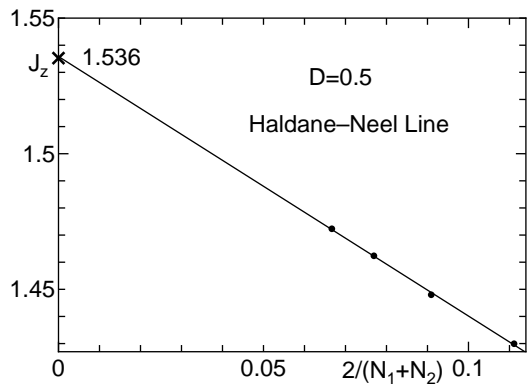}}
\caption{The extrapolation procedure of finite size critical point $J_{zc(\HN)}$ for $D=0.5$.}
\label{exphaneline}
\end{figure}

\begin{figure}[<h>]
\centerline{\includegraphics[width=70mm]{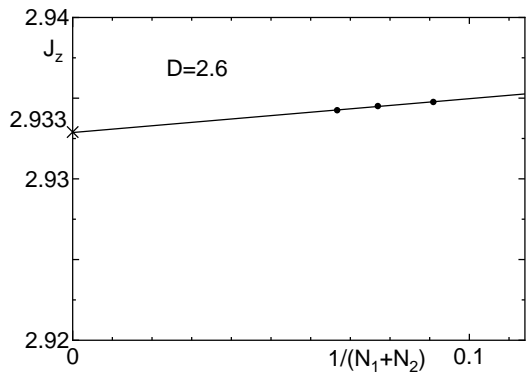}}
\caption{The extrapolation procedure of finite size $J_{zc(\HL)}$ for $D=2.6$ and $(N_{1},N_{2})$=(10,12),(12,14),(14,16).}
\label{exphanelar}
\end{figure}

Figure \ref{exphaneline} represents the extrapolation procedure of the Haldane-N\'eel transition point for $D=0.5$. The values of $J_{zc(\HN)}(N_1,N_2)$ for four pairs of system sizes $(N_{1},N_{2})=$(8,10), (10,12), (12,14) and (14,16) are represented by $\bullet$ in Figure \ref{exphaneline}. These values are extrapolated using the formula $J_{zc(\HN,\LN)}(N_{1},N_{2})=J_{zc(\HN,\LN)}(\infty)+2C_{1}/(N_{1}+N_{2})+4C_{2}/(N_{1}+N_{2})^2$ to obtain $J_{zc(\HN)}=1.536.$ The second term is necessary for small $D$ and $J_z$ for which the contribution from the irrelevant operators are not negligible for the present system size. Actually, as $D$ and $J_z$ increases the system size dependence becomes weak and for $2.4\simleq D \simleq3.0$, the system size dependence of the critical point is almost negligible as shown in Figure \ref{exphanelar}. The same procedure is carried out interchanging the roles of $J_{z}$ and $D$ if appropriate.

\begin{figure}[<h>]
\centerline{\includegraphics[width=70mm]{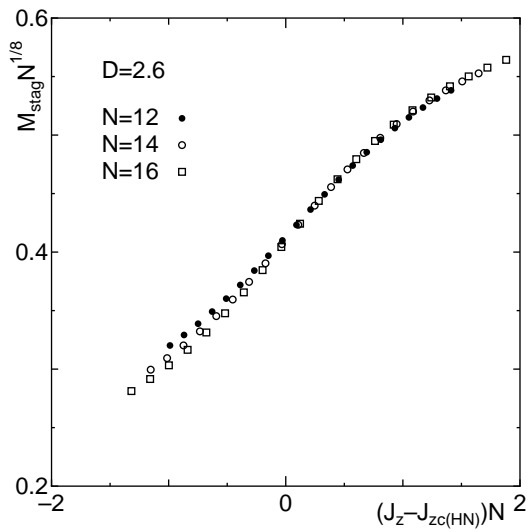}}
\caption{The finite size scaling plot of the staggered magnetization near the Haldane-N\'eel transition point for $N=$12,14 and 16. }
\label{fss}
\end{figure}

To check the Ising universality class, we also carried out the finite size scaling analysis of the staggered magnetization\cite{mn}.  The staggered magnetization operator is defined by $\hat{M}_{\stag}=\frac{1}{N}\sum_{i=1}^N S_i^z(-1)^i$. In finite size systems, the average $<\hat{M}_{\stag}>$ vanishes identically. Therefore, we calculate instead $M_{\stag}\equiv \sqrt{<\hat{M}_{\stag}^2>}$ by numerical diagnonalization with periodic boundary conditions for $N=8, 10, 12, 14$ and 16. The finite size scaling plot is shown in Figure \ref{fss} for $D=2.6$ with Ising exponent. It is clearly seen that the most data collapse onto a single curve for $J_z \simgeq J_{zc}$ (N\'eel side). On the Haldane side, the width of the Haldane phase is extremely small so that the data do not collapse well for $J_z < J_{zc}$ .

\begin{figure}[<h>]
\centerline{\includegraphics[width=70mm]{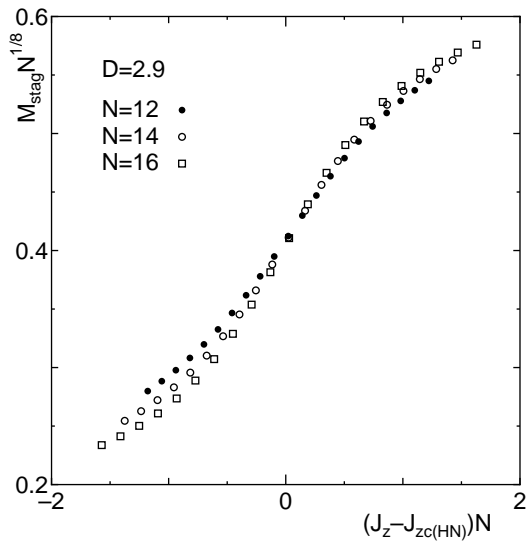}}
\caption{The finite size scaling plot of the staggered magnetization near the large-$D$-N\'eel transition point for $N=$12,14 and 16. }
\label{fss2}
\end{figure}

However these two specific features of the Ising critical line break down for  $D \simgeq 3.0$. Actually, the finite size scaling plot of $M_{\stag}$ assuming the Ising universality class fails to collapse onto a single line already for $D = 2.9$ as shown in Figure \ref{fss2}.  

\begin{figure}[<h>]
\centerline{\includegraphics[width=70mm]{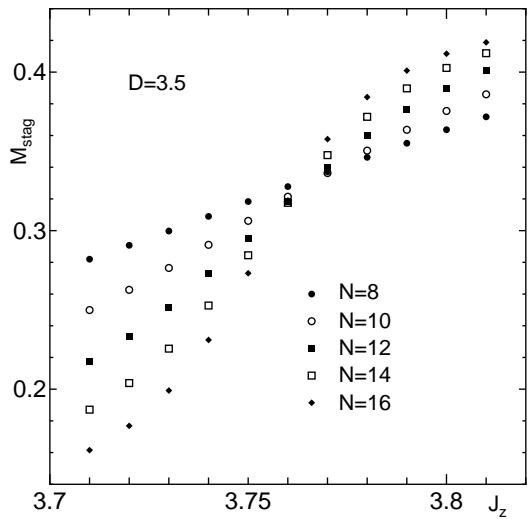}}
\caption{The $J_{z}$ dependence of the numerically obtained $M_{\stag}$ for $D=3.5$ for various values of $N$.}
\label{dneelcross}
\end{figure}

\begin{figure}[<h>]
\centerline{\includegraphics[width=70mm]{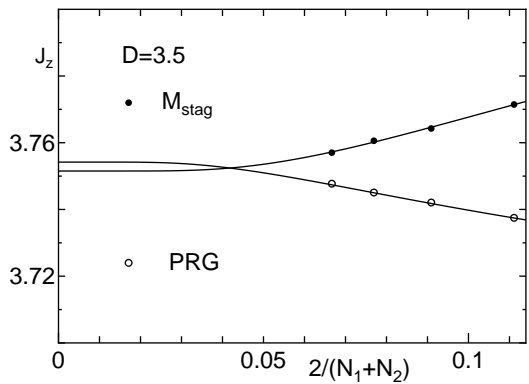}}
\caption{The extrapolation procedure of finite size $J_{zc(\LN)}$ to $N \rightarrow \infty$ by $J_{zc(\LN)}(N)=J_{zc(\LN)}(\infty)+C_{1}\exp(-(N_{1}+N_{2})/2\xi)$ for $D=3.5$. The critical points calculated from $M_{\stag}$ and PRG are represented by $\bullet$ and $\circ$, respectively.}
\label{expdneel}
\end{figure}

For $D > 3.0$, the system size dependence of the critical point again becomes large. Therefore we have also determined the critical point from the intersection point of $M_{\stag}$ for two successive system sizes $N_1=N$ and $N_2(=N+2)$ as shown in Figure \ref{dneelcross} for $D=3.5$. In this regime, if we assume the first order transition, the correlation length remains finite even at the transition point. Therefore the four intersections calculated by both methods for $(N_1,N_2)=$(8,10) (10,12) (12,14) and (14,16) are extrapolated to $N \rightarrow \infty$ by $J_{zc(\LN)}(N)=J_{zc(\LN)}(\infty)+C_{1}\exp(-(N_{1}+N_{2})/2\xi)$ as shown in Figure \ref{expdneel} for $D=3.5$.  The transition point in thermodynamic limit determined by the two methods coincide well as shown in Figure \ref{expdneel}. It should be noted that the transition points extrapolated by the power law do not coincide with each other in this regime. This confirms again that this transition is the first order transition.


\begin{figure}[<h>]
\centerline{\includegraphics[width=70mm]{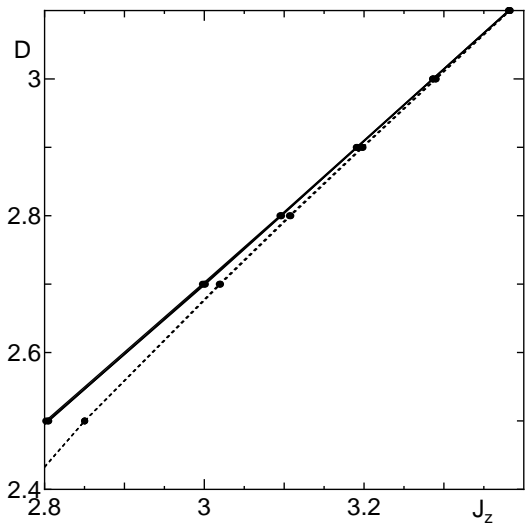}}
\caption{The enlarged figure of the phase boundary around the tricritical point. The solid (dotted) lines are the Haldane-large-$D$ (Haldane-N\'eel) critical lines extrapolated from various choices of finite size data.}
\label{tricri}
\end{figure}

The precise position of the Haldane-large-$D$-N\'eel tricritical point is difficult to determine. However, we estimate it from the point at which the Haldane-N\'eel critical line merges the Haldane-large-$D$ critical line. We carefully estimated the errors of both critical lines by trying the extrapolation to $N \rightarrow \infty$ choosing various sets of system sizes among $N=16, 14, 12, 10$ and $8$ as shown in Figure \ref{tricri}. From this figure, the two critical lines seem to merge around $(J_z, D) \sim (3.2,2.9)$.  As explained above, it is checked that the universality class clearly deviates from the Ising type around  $D=2.9$ while Ising universality class is confirmed around $D = 2.6$. Taking the whole situation described above into account, it is most likely that the tricritical point is located around $(J_z, D) \sim (3.2, 2.9)$ and the Haldane-N\'eel line is the Ising critical line and the large-$D$-N\'eel line is the first order line all the way down to the tricritical point.

\subsection{Large-$D$-XY and Haldane-XY transition line($J_{z}\le 0$ )}
From symmetry consideration, this transition is expected to be the Berezinskii-Kosterlitz-Thouless (BKT) transition. Because the BKT transition is a gapful-gapless transition, the critical points are difficult to determine. 
Following the procedure proposed by Nomura\cite{nomura,kn,kn2,kno}, the critical point is determined by the crossing point of the excitation energy of the lowest excitation $\Delta E_3$ with $M^z=4, P=1, k=0$ and  $\Delta E_0$  with $M^z=0,P=1, k=0$ where $k$ is the wave number of the excitation.

\begin{figure}[<h>]
\centerline{\includegraphics[width=70mm]{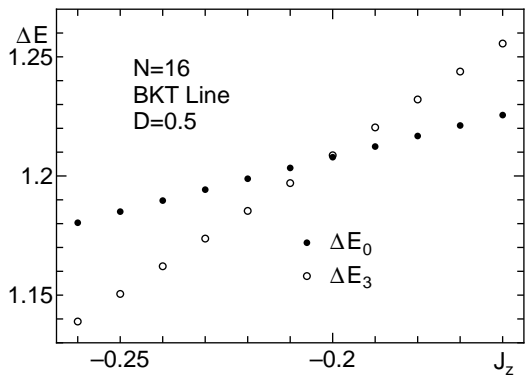}}
\caption{The $J_{z}$ dependence of the energy $\Delta E_{3}$ and $\Delta E_{0}$ represented by $\circ$ and $\bullet$, respectively, for $D=0.5$ and $N=16$.}
\label{xy-dcross}
\end{figure}

\begin{figure}[<h>]
\centerline{\includegraphics[width=70mm]{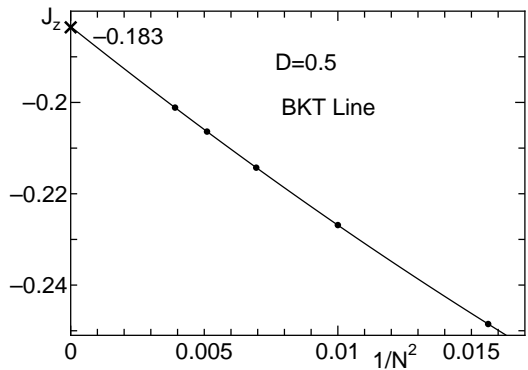}}
\caption{The extrapolation procedure of finite size $J_{zc(\LXY)}$ for $D=0.5$.}\label{expbktline}
\end{figure}

At the transition point these two energy levels cross as shown in Figure \ref{xy-dcross} for $N=16, D=0.5$. From the crossing point, we obtain the finite size large-$D$-XY transition point.  The BKT transition point for the infinite system can be obtained by extrapolating from $N=8, 10, 12, 14$ and 16 to $N \rightarrow \infty$ as $J_{zc(\LXY)}=-0.183$ as shown in Figure \ref{expbktline} for $D=0.5$. The extrapolated value is represented by $\times$.


The same procedure is carried out for the Haldane-XY transition line. From the results of numerical calculation, The critical points $J_{zc(\HXY)}$ are always equal to zero for the Haldane-XY transition. This can be understood in the following way for large negative $D$.  In this case, the spin states of the original model are restricted to  $\mid S^z=\pm 1>$ on each site. These states are identified with the effective $S=1/2$ spin states $\mid S_{\rm eff}^z=\pm 1/2>$. We use the perturbation method with respect to $1/D$ and $J_z$ to calculate the effective coupling between these effective $S=1/2$ spins. The effective Hamiltonian ${\cal H_{\rm {eff}}}$ is given by,
\begin{eqnarray}
\label {eff}
{\cal H_{\rm {eff}}}&=&\sum_{i=0}^{N} \Big[ \frac{1}{|D|}(S_{i}^{x}S_{i+1}^{x}+S_{i}^{y}S_{i+1}^{y}) \nonumber\\
&+&(\frac{1}{|D|}+4J_{z})S_{i}^{z}S_{i+1}^{z} \Big].
\end{eqnarray}
discarding the constant term. For $J_z=0$, this effective model becomes the isotropic antiferromagnetic Heisenberg model. It is exactly known that the XY-N\'eel transition takes place at the isotropic point for $S=1/2$ XXZ chain. Therefore the XY-N\'eel transition of the original model takes place at $J_z=0$ for large negative $D$.

\subsection{XY-ferromagnetic and large-$D$-ferromagnetic transition line ($J_z<0$)}

\begin{figure}[<h>]
\centerline{\includegraphics[width=70mm]{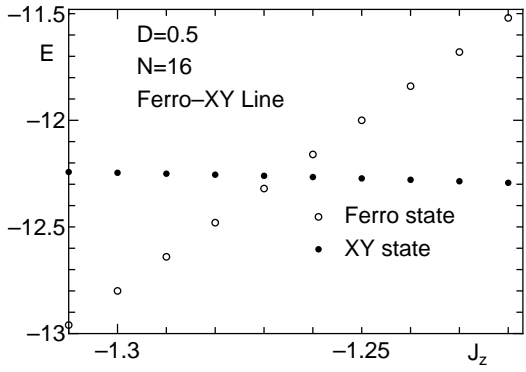}}
\caption{The $J_{z}$ dependence of the ground state energy of the XY phase and ferromagnetic phase is represented by $\bullet$ and $\circ$ , respectively, for $D=0.5$ and $N=16$.}
\label{fexycross}
\end{figure}

\begin{figure}[<h>]
\centerline{\includegraphics[width=70mm]{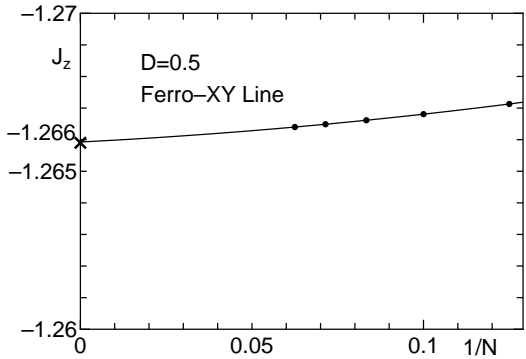}}
\caption{The extrapolation procedure of finite size $J_{zc(\XYF)}$ for $D=0.5$.}\label{expFeXYline}
\end{figure}

We can numerically verify that the ground state energy between the non-magnetic ground state with $M^z=0$ and fully polarized ground state with $M^z=N$ crosses at the XY-ferromagnetic transition line as shown in Figure \ref{fexycross} using the exact diagonalization for sizes $N=8,10,12,14$, and 16 with periodic boundary conditions. The partially polarized states have always higher energy. The crossing point is the finite size first order phase transition point $D_{c(\XYF)}(N)$ or $J_{zc(\XYF)}(N)$. We use $D_{c(\XYF)}(N)=D_{c(\XYF)}(\infty)+C_{1}N^{-2}+C_{2}N^{-4}$ or $J_{zc(\XYF)}(N)=J_{zc(\XYF)}(\infty)+C_{1}N^{-2}+C_{2}N^{-4}$ to extrapolate $D_{c(\XYF)}(N)$ or $J_{zc(\XYF)}(N)$ to $N \rightarrow \infty$ as shown in Figure \ref{expFeXYline}. The same procedure is carried out also for the large-$D$-ferromagnetic first order line. 


As explained in the preceding subsection, for $D \rightarrow -\infty$, this model can be described by the effective model (\ref{eff}). It is known that for $S=1/2$ XXZ chain, the XY-ferromagnetic transition takes place at the isotropic ferromagnetic point. Therefore the corresponding phase transition takes place at  $J_{z}=(2\mid D \mid)^{-1}$ for the original model. The numerically obtained transition line seems to approach this line for large enough negative $D$ as shown in Figure \ref{phase}.

\subsection{Transition line between two different XY phases ($D<0$ and $J_{z}<0$)}
\begin{figure}[<h>]
\centerline{\includegraphics[width=70mm]{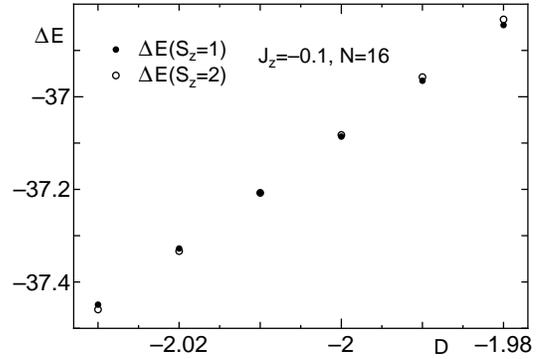}}
\caption{The $D$ dependence of the energy $E(M^z=\pm 2)$ and $E(M^z=\pm1)$ represented by $\circ$ and $\bullet$, respectively, for $J_z=-0.1$ and $N=16$.}
\label{2xy}
\end{figure}

\begin{figure}[<h>]
\centerline{\includegraphics[width=70mm]{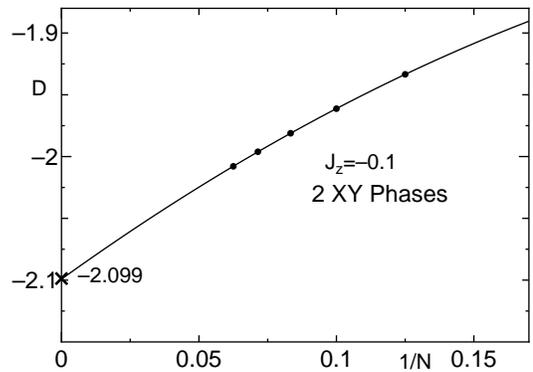}}
\caption{The extrapolation procedure of finite size $D_{c(\XYXY)}$ for $J_z=-0.1$.}
\label{expxy}
\end{figure}

Within the XY phase, there are 2 different types of phases as predicted by Schulz\cite{hj}. For large negative $D$, we find the lowest excited state with the excitation energy of the order of $1/N$ has quantum number $M^z=\pm 2$ that corresponds to the $M^z=\pm 1$ excitation in the effective Hamiltonian (\ref{eff}). This phase corresponds to the XY phase of the effective model. In this phase, the $M^z=\pm 1$ excitation can be only excited by forming the local $\mid 0>$ state that has the finite energy gap of the order of $\mid D \mid$. With decreasing $\mid D \mid$, the $M^z= \pm 1$ excitation becomes the lowest with excitation energy of the order of $1/N$. This phase is continuously connected with the XY phase of the $S=1$ XXZ model with $D=0$. Corresponding to the change of the quantum number of the lowest excitation, these two phases have different type of quasi-long range order. In the XY phase with large negative $D$ (XY2 phase), the correlation functions $<S^{x2}_iS^{x2}_j>$ and $<S^{y2}_iS^{y2}_j>$ decay with a power law dependence while $<S^{x}_iS^{x}_j>$ and $<S^{y}_iS^{y}_j>$ decay exponentially. On the other hand, in the XY phase with small negative $D$ (XY1 phase), the correlation  functions  $<S^{x}_iS^{x}_j>$ and $<S^{y}_iS^{y}_j>$ decay with a power law. Therefore they can be regarded as two different phases. The level crossing point of the $M^z=\pm 1$ excitation and $M^z= \pm 2$ excitation is the critical point between these two XY phases. An example is shown in Figure \ref{2xy} for $J_z=-0.1$ and $N=16$. The $M^z= \pm 1$ gap and the $M^z= \pm 2$ gap are shown by  $\bullet$ and $\circ$, respectively. The value of $D$ on the intersection point is $D_{c(\XYXY)}(N)=-2.008$. We use $D_{c(\XYXY)}(N)=D_{c(\XYXY)}+C_1N^{-1}+C_2N^{-2}$  to extrapolate $D_{c(\XYXY)}$ to $N \rightarrow \infty$ for $N=8,10,12,14,$ and 16 as shown in Figure \ref{expxy}. The same procedure is carried out appropriately interchanging the roles of $J_z$ and $D$.

\section{Summary and Discussion}

The ground state phase diagram of a spin-1 XXZ chain with uniaxial single-ion-type anisotropy is determined accurately by analyzing the numerical diagonalization data using the level spectroscopy, conformal field theory analysis, the phenomenological renormalization group and finite size scaling.

 Most parts of the phase diagram is determined accurately and the universality class of most critical lines are obvious from symmetry consideration. The phase transition between the large-$D$ phase and N\'{e}el phase is very likely to be a first order transition as expected from the consideration of the large $D$ limit, although we have no final proof that it is so  all the way down to the tricritical point. In this context, it is of great interest how the first order or Ising type transition line splits into a Gaussian (large-$D$-Haldane) and Ising (Haldane-N\'eel) lines. 
 
 Related to this problem, the precise position of the tricritical point remained ambiguous. We have determined it from the point where the numerically obtained large-$D$-Haldane critical point and large-$D$-Ising critical point merge and the finite size scaling analysis of the staggered magnetization also supports this estimation. However it is difficult to determine this point accurately by numerical analysis. Further analytical insight into the properties of the tricritical point is necessary to elucidate this issue.

One of the authors (KH) thanks M. Nakamura, K. Nomura, K. Okamoto and S. J. Qin for enlightening discussion. The computation in this work has been done using the facilities of the Supercomputer Center, Institute for Solid State Physics, University of Tokyo, the Information Processing Center and Department of Physics of Saitama University.  The diagonalization program is based on the TITPACK ver.2 coded by H. Nishimori.  This work is supported by a research grant from the Natural Science and Engineering Research Council of Canada (NSERC) and a Grant-in-Aid for Scientific Research from the Ministry of Education, Science, Sports and Culture of Japan.


\begin{thebibliography}{99}
\bibitem{bh}
H. Bethe: Z. Phys. {\bf 71} 205 (1931).
\bibitem{fd}
F. D. M. Haldane: Phys. Lett. {\bf 93A} (1983); Phys. Rev. Lett. {\bf 50} 1153 (1983).
\bibitem{nomura}
K. Nomura: J. Phys. A: Math. Gen. {\bf 28} 5451 (1995).
\bibitem{hj}
H-J. Schulz: Phys. Rev. {\bf B34} 6372 (1986).
\bibitem{md}
M. den Nijs and K. Rommelse: Phys. Rev. {\bf B40} 4709 (1989).
\bibitem{kno}
A. Kitazawa, K. Nomura and K. Okamoto: Phys. Rev. Lett. {\bf 76} 4038 (1996).
\bibitem{kn}
A. Kitazawa and K. Nomura: J. Phys. Soc. Jpn. {\bf 66} 3944 (1997).
\bibitem{kn2}
A. Kitazawa and K. Nomura: J. Phys. Soc. Jpn. {\bf 66} 3379 (1997).
\bibitem{ak} 
A. Kitazawa: J. Phys. A. Math. Gen. {\bf 28} L285 (1997).
\bibitem{ht}
H. Tasaki: Phys. Rev. Lett {\bf 66} 798 (1991).
\bibitem{tt}
T. Tonegawa, T. Nakao and M. Kaburagi: J. Phys. Soc. Jpn. {\bf 65} 3317 (1996).
\bibitem{ca}
J. L. Cardy: J. Phys. {\bf A17} L385 (1984).
\bibitem{HW} H. W. J. Bl\"ote, J. L. Cardy and M. P. Nightingale: Phys. Rev. Lett. {\bf 56} 742 (1986).
\bibitem{IA} I. Affleck: Phys. Rev. Lett. {\bf 56} 746 (1986); Nucl. Phys. B {\bf 270 [FS16]} 186 (1986).
\bibitem{mn} M. N. Barber: {\it Phase Transitions and Critical Phenomena 8}, ed. C. Domb and J. L. Lebowitz (Academic Press, London, New York, 1983) 146.
\end{thebibliography}
\end{document}